\documentclass[12pt]{article}
\usepackage[german,english]{babel}   
\usepackage{graphics}
\usepackage{graphicx}
\newcommand{\rt}{$\langle r \rangle_{(2)}$}
\newcommand{\rp}{$\langle r^2 \rangle^{1/2}_p$}

\newcommand{\dt}{d^3}
\newcommand{\aei}{$\alpha^8$}
\newcommand{\be}{\begin{eqnarray}}
\newcommand{\ee}{\end{eqnarray}}
\newcommand{\bes}{\begin{eqnarray*}}
\newcommand{\ees}{\end{eqnarray*}}

\begin{document}
\vspace*{-0.6in}
\hfill \fbox{\parbox[t]{1.12in}{LA-UR-03-7323}}
\vspace*{0.0in}

\vspace*{0.2cm}
\begin{center}
{\bf \LARGE Zemach Moments for Hydrogen and Deuterium}\\
\vspace{1cm} 
{\bf \large
J. L. Friar$^1$ and Ingo  Sick$^2$}\\
\vspace{0.5cm}
{\small \sl 
$^1$Theoretical Division, Los Alamos National Laboratory, \\ Los Alamos, New
Mexico 87545 \\
$^2$Dept. f\"ur Physik und Astronomie,
Universit\"at Basel  \\ CH-4056 Basel, Switzerland \\}
\end{center}
\vspace{1.2cm}                 
\normalsize
\begin{center} Abstract \end{center}
\small
\begin{minipage}[t]{1.0cm}
\hfill
\end{minipage}
\begin{minipage}[t]{12cm}
We determine the Zemach moments of hydrogen and deuterium for the first time
using only the world data on elastic electron-proton and electron-deuteron
scattering. Such moments are required for the calculation of the nuclear
corrections to the hyperfine structure of these hydrogenic atoms. We compare the
resulting HFS predictions to the available high-precision data and provide an
estimate of the size of the nuclear polarization corrections necessary to
produce agreement between experimental HFS and theoretical calculations.
\end{minipage} \\
 \normalsize
\newpage  {\bf Introduction.} \hspace{0.3cm} 
Nuclear and atomic size scales differ by nearly five orders of magnitude, which
makes nuclear corrections to atomic energy levels very tiny.  The exceptional
precision of modern microwave and optical measurements of atomic level spacings
nevertheless makes these nuclear effects significant.  The recent
measurement\cite{Niering00} of the 1S-2S interval in {\em hydrogen} to an
unprecedented accuracy of 2 parts in $10^{14}$ is affected in the tenth
significant figure by the finite size of the proton, and this explains the
strong interest of the atomic physics community in the value of the proton's
$r.m.s.$ charge radius\cite{Eides01,Friar02}. The situation for the {\em deuterium
atom} is similar. 

The cloudy history of experimental values for the {\em proton} radius has recently been
clarified by a comprehensive analysis\cite{Sick03} of all the world's
electron-proton scattering data.  That work separated the charge and magnetic
scattering, incorporated (significant) Coulomb corrections\cite{Sick96b}, and
carefully treated systematic (as well as statistical) uncertainties.  The 
resulting value 
of
\rp = 0.895(18)fm is significantly higher than
most older values, and is in line with values obtained from analyses of the Lamb
shift in atomic hydrogen\cite{Melnikov00,Pachucki01,Eides01}.   With 
the inclusion of sufficiently
large QED corrections of order \aei \ \cite{Pachucki01,Jentschura03}
 it should soon be possible to extract
values of the proton radius from the hydrogen 1S-2S interval that are an order
of magnitude more precise than that of Ref.\cite{Sick03}.  Even more precise 
values of the proton radius might result from the ongoing PSI experiment to
measure the Lamb shift in muonic hydrogen\cite{Taqqu99}.

The other area where the finite size (or internal structure) of the proton plays
a significant role is the hyperfine structure of the nS levels of hydrogen.  A
combined analysis of various experiments measuring the 1S hyperfine splitting is
given in Ref.\cite{Karshenboim99}, which advocates a value
$$
   \Delta E_{\rm hfs}^{\rm exp} = 1\, 420\, 405.751\, 768(1)\, {\rm kHz}
$$
that has an accuracy of better than one part in $10^{12}$.  The size of the 
proton affects the sixth significant figure.   It is the magnetic nature of
the hyperfine interaction that leads to this enhanced sensitivity to nuclear
({\em i.e.} short-range) properties. Hyperfine structure is much more sensitive than
the Lamb shift to the high-frequency components of the electromagnetic
interactions that bind atoms.

The hyperfine mechanisms are traditionally divided into three categories:  pure
QED, recoil, and nuclear size and structure.  The pure QED contributions are
listed and discussed in Refs.\cite{Eides01} and \cite{Karshenboim02}, and
uncalculated terms are expected to be significantly smaller that 1 ppm relative
to the Fermi hyperfine splitting.  Recoil
(or nuclear-mass-dependent) terms\cite{Bodwin88} are usually lumped together
with the nuclear corrections.  Because of the sensitivity to high-frequency or
short-range interaction terms, QED for systems with a fundamental anomalous
magnetic moment is not renormalizable, and as a result some recoil corrections
are divergent without nuclear form factors. Although nominally the same size as
the static nuclear-size correction ({\em i.e.} the Zemach correction\cite{Zemach56}
of about $\mbox{$-$}$40 ppm relative to the Fermi hyperfine splitting and
discussed below), the leading-order recoil correction is substantially smaller
(about 5 ppm) and has only a logarithmic dependence on nuclear structure, which
produces a rather smaller uncertainty, as well\cite{Eides01}.

Although one would naively expect (in analogy with the Lamb shift) that for HFS 
the leading-order nuclear-size correction is given by a simple average over
the nuclear magnetic density, this is not the case. The leading-order nuclear
size and structure corrections for hyperfine splittings actually arise from
two-photon-exchange diagrams and are usually divided into contributions from
elastic and inelastic nuclear intermediate states (plus appropriate nuclear
seagull terms).  The inelastic contributions (polarization corrections) can be
expressed as integrals over the spin-dependent electron-scattering structure
functions\cite{Hughes83}, $g_1$ and $g_2$, and are very difficult to calculate 
reliably\cite{Drell67,Iddings69,Faustov02}. An upper limit of $\pm$ 4 ppm 
exists\cite{Hughes83}, although calculations using resonance models and 
existing data find smaller values (typically 1-2 ppm)\cite{Faustov02}. The 
nuclear-structure-dependent corrections for the proton are therefore 
completely dominated by the elastic part, and that is the purview of this
work.

For the {\em deuteron} the HFS is also known with excellent accuracy 
\cite{Karshenboim99}
$$
   \Delta E_{\rm hfs}^{\rm exp} =  327\, 384.352\, 522(2)\, {\rm kHz} 
$$
where nuclear effects amount to about 138 ppm. 
The  deuteron presents a very different theoretical problem based on very 
different scales. Because the deuteron is so loosely bound it is very 
susceptible to breakup reactions. The deuteron Zemach correction is about --100 
ppm (see below), leaving an inelastic contribution of about 240 ppm.
The bulk of the nuclear-size corrections 
to hyperfine structure is therefore generated by inelastic 
contributions\cite{Milshtein96}, although the elastic contribution is clearly
important, particularly in view of the cancellation.
It is extremely valuable for any theorists attempting to perform 
these calculations to be able to judge the quality of their work by comparison 
to appropriate experimental results. For this reason we also determine  the 
Zemach moment for the deuterium atom, even though it is not the dominant part of
the nuclear contribution.

{\bf Zemach moments.~~} 
The bulk of the electron-nucleus magnetic interaction is short ranged and
confined to the vicinity of the nucleus.  This is also the only region of the
electron's wave function that is significantly affected by the nuclear charge
distribution, and the leading-order size effect was shown by 
Zemach\cite{Zemach56} to depend on the product of the proton's elastic charge 
and magnetic form factors (a convolution in configuration space) in the form
\be
\Delta E_{\rm Zemach} = -2\, Z\, \alpha\, m\, \langle r \rangle_{(2)}\, 
E_{\rm F}
\ee
\be
\langle r \rangle_{(2)} & = & \int \dt r\, r \int \dt r^{\prime} \rho_{\rm ch} 
(|{\bf r - r^{\prime}}|) \rho_{\rm mag} (r^{\prime})  \nonumber \\ & = & 
-\frac{4}{\pi}
\int_0^{\infty} \frac{d q}{q^2} (G_{\rm E} (q^2) G_{\rm M} (q^2) - 1) 
\label{gegm} \ee
where $E_{\rm F}$ is the Fermi hyperfine splitting, $m$ is the electron mass,
$Z$ is the nuclear charge, $\alpha$ is the fine-structure constant, $G_{\rm E}
(q^2)$ $(G_{\rm M})$ is the charge (magnetic) form factor depending only on the
momentum transfer (squared), $q^2 > 0$. The subtraction term ($\mbox{$-$}$1) is
necessary in order to avoid double counting the nuclear  charge and magnetic 
moment. The convoluted
configuration-space densities $\rho_{\rm ch}$ and $\rho_{\rm mag}$ are simple
Fourier transforms of the form factors, both of which are normalized to 1 at
$q^2 = 0$.

The same problems that have plagued extraction of the proton's and deuteron's
charge radii 
have also complicated the calculation of $\langle r \rangle_{(2)}$ via the 
momentum-space integral above.  Most calculations have relied on a common dipole
shape for both charge and magnetic form factors and have presented results based
on various values of the single parameter in such shapes.  That parameter also
determines the value of the (common) mean-square radius, which historically has
had well-scattered values. Results for the proton 
typically\cite{Eides01} have been in the 
vicinity of $\langle r \rangle_{(2)} \sim 1.0$ fm and $\Delta E_{\rm Zemach} 
/E_{\rm F} \sim -40$ ppm. We will use the electron-scattering data themselves, 
together with well-tested techniques for extracting the form factors, to 
evaluate $\langle r \rangle_{(2)}$.

{\bf Determination of} $\mathbf{\langle r \rangle_{(2)}}$ \hspace{0.3cm}
In previous papers \cite{Sick96b,Sick03,Sick01} we have described our analysis of the
{\em world} data on $e-p$ and $e-d$ scattering. Here, we use these results to 
determine the Zemach moments. 

The proton cross sections up to the maximum momentum transfer 
$q_{max}=4fm^{-1}$ have
been fit with a 5-parameter continued fraction expansion for both $G_{E}(q)$ and
$G_{M}(q)$.  The deuteron data up to $q_{max}=8fm^{-1}$ have been fit
with a 10-parameter SOG parameterization for the electric monopole (C0), 
magnetic dipole (M1) and  electric quadrupole (C2) form factors.  The references to the cross 
sections and polarization data included are listed in refs.\cite{Sick96b,Sick03}. 
In the fits, the Coulomb distortion of the electron waves,
neglected in all work before \cite{Sick96b}, has been included.

The separation of longitudinal (charge)  and transverse (magnetization)
contributions to the (e,e) cross sections is automatically performed during the
fit of the cross section data. For the case of the deuteron, the separation 
of monopole and quadrupole
contributions is also achieved as all the available polarization data are
included in the data set.

An important feature of these fits is the fact that charge and magnetic form
factors are {\em simultaneously} fit to the available data set. The error 
matrix of
the fit then contains all the correlations between the two (three) form factors, 
resulting from the fact that the observed cross sections depend on a linear 
combination of charge and magnetic form factors. These correlations obviously 
are important when computing the uncertainty in the
Zemach integral, eq.\ref{gegm}.  As the charge/magnetic- (L/T)-separation 
leads to an {\em anti}correlation
between $G_E$ and $G_M$ and the Zemach moment depends on $G_E \cdot G_M$, 
the Zemach moment actually to some degree can be determined better than 
quantities depending only on $G_E$ or $G_M$, such as, {\em e.g.} the
$rms$-radii. In order to calculate the statistical uncertainty
of \rt , we use the corresponding error matrix. 

The systematic uncertainties of the data, mainly errors in the absolute 
normalization of the cross sections, are also affecting the uncertainty of the
Zemach moments. We determine this uncertainty by changing the individual data
sets by their quoted systematic uncertainty, refitting the form factors 
 and adding quadratically the resulting changes of the Zemach moments.

From the parameterization of the data we determine the contribution to the 
integral eq.\ref{gegm} up to $q=q_{max}$. We add the contribution from $q_{max}$
to $q=\infty$ using a dipole form factor. We have verified that, due to the
smallness of the product $G_E G_M$  at large $q$, more realistic values 
for $G(q)$ do not make a significant difference.
 
 The results are listed in Table \ref{zema}. 
The error bar given covers  both the statistical and the systematic
uncertainties of the data.

\begin{table}[htb] 
\begin{center}
\begin{tabular}{l|l}
~ Nucleus ~~~& ~~~~~ Zemach-moment ~~~ \\[2mm]
\hline
\rule[0mm]{0mm}{6mm} Proton  & ~~~ 1.086$ \pm$0.012 $fm$ \\
\rule[0mm]{0mm}{6mm} Deuteron& ~~~ 2.593$ \pm$0.016 $fm$ \\[2mm]
\hline
\end{tabular} \\
\parbox{6cm}{\caption[]{
\rt \ from (e,e) data. 
}\label{zema}} 
\end{center}
\end{table}

For comparison, we have also calculated the moments for some standard
parameterizations of the form factors. When using for the proton the 
conventional dipole form
factor, one finds 1.023$fm$, while the Hoehler 8.2 fit \cite{Hoehler76}
gives 1.038$fm$. For the deuteron, a
dipole parameterization, with the best-fit charge $rms$-radius \cite{Sick96b}
determining the scale parameter, gives 2.679$fm$, while the 
zero-range-approximation model for the deuteron of ref. \cite{Milshtein96} 
yields 1.708$fm$. 

If one calculates the deuteron charge and magnetic form factors
using the impulse approximation and the AV18 potential model\cite{Wiringa95}, 
together with a dipole nucleon form factor adjusted to give a proton Zemach 
moment of 1.086 $fm$, one finds a deuteron Zemach moment of 2.656 $fm$, which 
is about 2\% too high. This illustrates an important point: the deuteron's 
Zemach moment depends on a wide range of physics contributing to the charge and 
magnetic form factors, from the potential 
model used to calculate the wave functions, to possible meson-exchange currents 
(ignored in the impulse approximation) and relativistic corrections, and to the 
nucleon form factors themselves. Although the details of these ingredients are
expected to produce only a small overall effect (a few percent, at most), the 
precision of our result for the deuteron pins down the size of the defect.

These values show that the Zemach moments are quite sensitive to the
$q$-dependence of the form factors employed; they do not only depend on the
$rms$-radii. The values given also show that the Zemach moments do depend 
appreciably on the {\em difference} between the charge and the magnetic form 
factors. 

We note that the uncertainties on the Zemach moments are in part smaller than
what could be expected from  the uncertainties of the corresponding
$rms$-radii \cite{Sick03,Sick01}. 
We attribute this to two distinct reasons: (i) the  anti-correlation of $G_E$
and  $G_M$ mentioned above. (ii) Sensitivity studies of the Zemach integral show that \rt~
depends on the form factors $G(q)$ up to $q \sim 3 fm^{-1}$, and not only on 
the low-$q$ properties (radii); at these larger $q$'s the finite-size effect 
in the form
factors is bigger, with a correspondingly reduced importance of the systematic
uncertainties of the data that dominate the uncertainty in the radii.

{\bf Hyperfine splitting. \  }
The hyperfine splitting of the 1S level of {\em hydrogen} was calculated using the
fundamental constants from the 1998 CODATA evaluation\cite{Mohr00} and the 
QED and recoil
corrections listed in Ref.\cite{Eides01} (see also \cite{Karshenboim02}). 
We note that only the leading-order recoil correction calculated in 
Ref.\cite{Bodwin88} is significant, because the sum of calculated recoil and 
structure terms of  sub-leading order \cite{Karshenboim97} cancel almost
perfectly, leaving a negligible residue\cite{Eides01}. Our result for the 
Fermi splitting is 1\,418\,840.1 kHz, while adding in the QED and Breit 
corrections  leads 
to 1\,420\,452.0 kHz, both results the same as that of Table XVIII of 
Ref.\cite{Eides01}.
Subtracting the experimental result from this theoretical result leads to a 
residue of 32.6 ppm of the Fermi splitting, which must accommodate all 
recoil and  strong-interaction
effects. The leading-order recoil contribution is 5.22 ppm, leaving 37.8 ppm 
for the sum of Zemach plus polarization corrections. Our Zemach moment 
correction  of --41.0(5) ppm
leaves the experimental result 3.2(5) ppm larger than theory without 
polarization. A recent calculation of the polarization 
correction found 1.4(6) ppm, which has the appropriate sign to complement 
our result,  but leaves the experimental result larger than theory by
1.8(8) ppm, which is about two standard deviations from zero. This difference 
accounts for the smaller value of \rt deduced by Ref. \cite{Dupays03}, which
assumed that the polarization corrections of Ref.\cite{Faustov02} are 
numerically accurate. Theoretical
error estimates are highly subjective and the polarization corrections (or 
the recoil corrections) may be somewhat more uncertain than believed. 
Our residue of  3.2(5) ppm without
polarization is within the upper limit for polarization corrections. Given 
the precision of our result for the Zemach moment, more attention to the 
polarization correction would be welcome.

The {\em deuterium} HFS involves significantly larger nuclear corrections. The QED
contribution is 337\,339.1 kHz, leaving a 138 ppm residue for nuclear plus recoil
corrections. Our deuteron Zemach moment in Table 1 generates a --98.0(6) ppm
contribution, so that polarization and recoil must contribute about 236 ppm. 
The reader is directed to Refs.\cite{Milshtein96} and \cite{Khriplovich03} 
for a
discussion of this polarization contribution which is very  difficult to
calculate.

{\bf Conclusion.~~} 
In this paper we have determined directly from the electron scattering data 
the Zemach moments for the proton and the deuteron. In particular for the
proton, this allows a much more precise comparison between the theoretical and
experimental HFS; the present status is agreement within 1.8(8)ppm, the main
source of uncertainty  being the proton polarizability. 
%


\end{document}